\documentclass[11pt,twoside]{article}
\usepackage{natbib}
\usepackage{amsmath,amssymb}
\usepackage{asp2010}

\resetcounters

\begin{document}

\title{Modeling scattering polarization for probing solar magnetism}
\author{J.~Trujillo~Bueno$^{1,2}$
\affil{$^1$Instituto de Astrof\'{\i}sica de Canarias, 
E-38205 La Laguna, Tenerife, Spain}
\affil{$^2$ Consejo Superior de Investigaciones Cient\'\i ficas, Spain}}

\begin{abstract}
This paper considers the problem of modeling the light polarization that emerges from an  
astrophysical plasma composed of atoms whose excitation state is significantly influenced by the anisotropy of the incident radiation field. In particular, it highlights how radiative transfer simulations  
in three-dimensional models of the ``quiet" solar atmosphere
may help us to probe its thermal and magnetic structure, from the near equilibrium photosphere to the highly non-equilibrium upper chromosphere. The paper finishes with predictions concerning 
the amplitudes and magnetic sensitivities  
of the linear polarization signals produced by scattering processes in two transition region lines, which should encourage us to develop UV polarimeters for sounding rockets and space telescopes with the aim of opening up a new diagnostic window in astrophysics. 
\end{abstract}

\section{Introduction}

Consider a line of sight (LOS) that makes an angle $\theta=45^{\circ}$ with respect to the solar radius through the observed point (i.e., ${\,}{\mu}={\rm cos}\,{\theta}=0.707$) and let us make the following virtual measurements of the emergent Stokes profiles in a ``non-LTE line". Assume that the observed plasma structure is 
strongly magnetized (with magnetic strength $B=1200$ G), and consider two different magnetic field orientations: (1) vertical and (2) horizontal (see Fig. 1). The dotted lines of Fig. 2 show the emergent Stokes profiles produced by the Zeeman effect, which for this LOS turn out to be identical for both field orientations because of the well-known azimuth ambiguity of the Zeeman effect. However, the Stokes profiles produced by the Sun are very different. Actually, the Stokes profiles corresponding to the vertical field case are given by the solid lines of Fig. 2, while those corresponding to the horizontal field case are given by the dashed lines. 

\begin{figure}[!h]
\begin{center}
\includegraphics[width=5.0 cm]{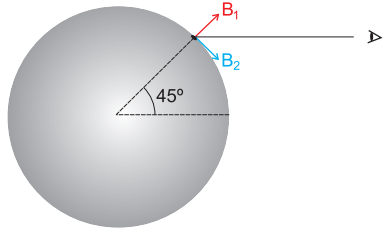}
\end{center}
\caption{The LOS and the magnetic field orientations considered in Figs. 2 and 3.}
\label{trujillo_fig:sun}
\end{figure}

\begin{figure}[!t] 
\begin{center}
\includegraphics[width=11.0 cm]{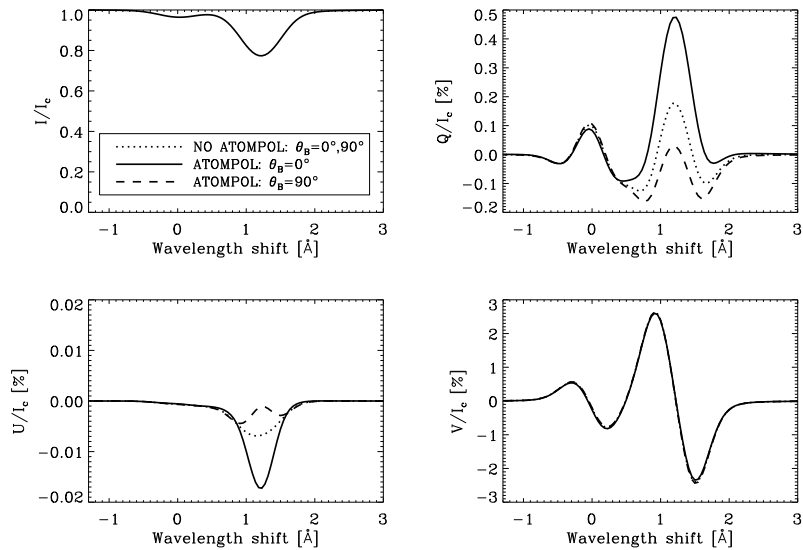}
\end{center}
\caption{Emergent Stokes profiles of the He {\sc i}~10830~\AA\ triplet for $B=1200$ G, calculated with 
the spectral synthesis option of the code HAZEL (from HAnle and ZEeman Light) developed by \cite{jtb_hazel}.
The ``ATOMPOL'' profiles have been obtained taking into account the joint action of atomic level polarization and the Hanle and Zeeman effects, while the ``NO ATOMPOL'' profiles are produced by the Zeeman effect alone. 
The reference direction for positive Stokes $Q$ is the perpendicular to the plane formed by the
solar radius vector through the observed point and the LOS. 
}
\label{trujillo_fig:b1200}
\end{figure}

We may thus note that

\begin{itemize}

\item The polarization of the Zeeman effect only depends on the orientation of the magnetic field with respect to the LOS.

\item The polarization produced by the Sun depends on both the orientation of the magnetic field with respect to the LOS and on its inclination with respect to the local vertical. 

\end{itemize}

\begin{figure}[!t]
\begin{center}
\includegraphics[width=11.0 cm]{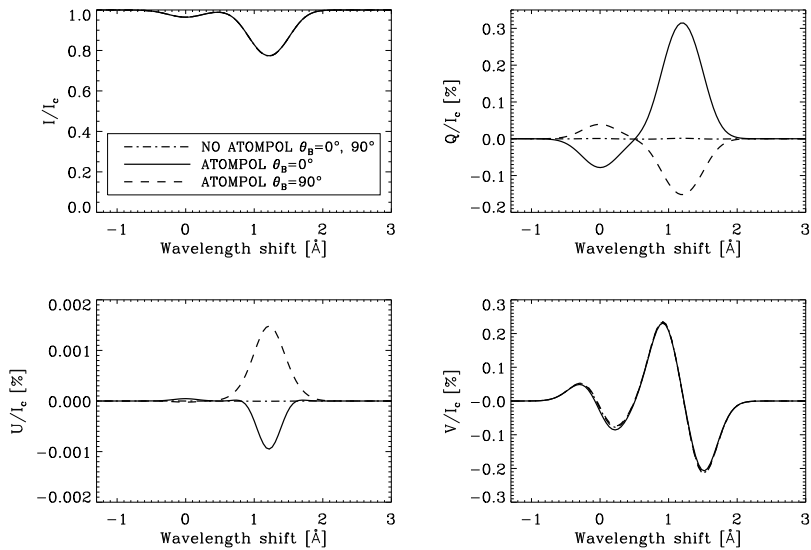}
\end{center}
\caption{As in Fig. 2, but for $B=100$ G.
}
\label{trujillo_fig:b100}
\end{figure}

Why does the Sun behave this way? Because we are dealing here with an astrophysical plasma composed of atoms whose excitation state is significantly influenced by the anisotropy of the pumping radiation field. The ensuing  radiative transitions produce population imbalances and quantum coherences between the magnetic sublevels pertaining to the lower and/or upper levels of the spectral line under consideration (i.e., atomic level polarization), in such a way that the populations of magnetic substates with different values of $|M|$ are different. This is termed atomic level alignment. The radiatively induced polarization of the atomic levels produces selective emission and/or selective absorption of polarization components, which give rise to linear polarization in the emergent spectral line radiation even in the absence of magnetic fields (e.g., Trujillo Bueno 2001, for a detailed review). Via the Hanle effect the atomic level polarization is sensitive to the presence of a magnetic field inclined with respect to the symmetry axis of the incident radiation field, so that it is natural to find in Fig. 2 very different results for the vertical and horizontal magnetic field cases.

What kind of Stokes profiles would we observe if the magnetic strength were much lower (say, $B=100$ G) ? As seen in Fig. 3, if $B=100$ G the circular polarization of the He {\sc i} 10830 \AA\ triplet is still dominated by the longitudinal Zeeman effect, but the linear polarization is fully controlled by atomic level polarization and the Hanle effect with no  significant influence of the transverse Zeeman effect. Interestingly, this is precisely the situation we find in quiet regions of the solar atmosphere for many spectral lines, not only for the He {\sc i} 10830 \AA\ triplet considered in Figs. 2 and 3. An observational example is shown in Fig. 4, which corresponds to high-sensitivity spectropolarimetric measurements of the Ca {\sc ii} IR triplet obtained in collaboration with R. Ramelli (IRSOL) using the Z\"urich Imaging Polarimeter (ZIMPOL) at the THEMIS telescope. This is because for weak magnetic fields (with $B{\lesssim}100$ G) the Zeeman splitting of the $\sigma$ and $\pi$ components of many spectral lines is negligible compared with the spectral line width.

\begin{figure}[!t]
\begin{center}
\includegraphics[width=7.6 cm]{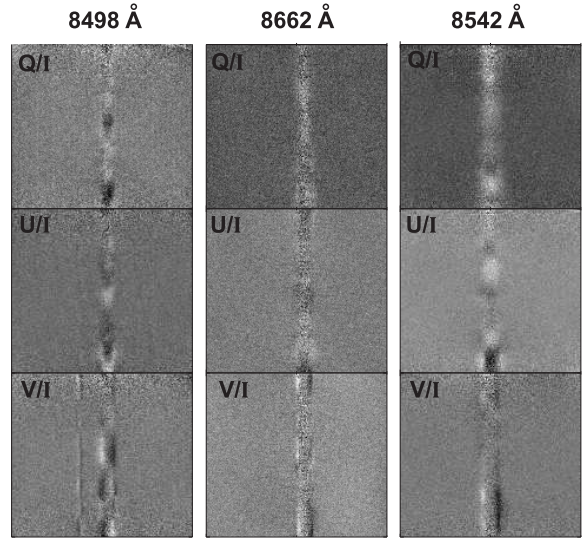}
\end{center}
\caption{An example of our recent spectropolarimetric observations of the Ca {\sc ii} IR triplet in a quiet region close to the solar limb, using ZIMPOL at the telescope THEMIS of the Observatorio del Teide. Note that the sign of the $U/I$ signals fluctuates along the spatial direction of the spectrograph's slit, which suggests spatial variations in the azimuth of the magnetic field in the quiet solar chromosphere. The reference direction for Stokes $Q$ is as in Fig. 2. From Trujillo Bueno et al. (2011).}
\label{trujillo_fig:zimpol-calcium}
\end{figure}

This paper highlights how radiative transfer modeling of the 
scattering polarization observed in the ``quiet" solar atmosphere 
allows us to explore its thermal and magnetic structure, but
it does not provide information on numerical methods. Such information can be found 
in several recent reviews. For the case of partial frequency redistribution (PRD), but assuming a two-level model atom with no lower-level polarization, see \cite{jtb_nagendra-spw5}. For the case of realistic multilevel atoms with the possibility of atomic level polarization in all the levels, but assuming complete frequency redistribution (CRD), see \cite{jtb_trujillo-tubinga}. Unfortunately, a general PRD theory for modeling scattering line polarization with multilevel atoms is not yet available \citep[e.g.,][]{jtb_belluzzi-spw6}. Nevertheless, the CRD theory described in the monograph by \cite{jtb_landi-landolfi} is suitable for modeling the polarization observed in the Sun's continuous radiation (see Section 2.1) and in many diagnostically important spectral lines, such as the He {\sc i} 10830 \AA\ and D$_3$ multiplets \citep[e.g.,][]{jtb_casini-landi-07}, the Sr {\sc i} line at 4607 \AA\ (see Sections 2.2 and 2.3), the lines of ionized cerium \citep[][]{jtb_manso-cerium}, the CN lines \citep[e.g.,][]{jtb_asensio-trujillo-spw3}, 
the Li {\sc i} 6708 \AA\ line \citep[][]{jtb_belluzzi-lithium}, the Mg {\sc i} $b$-lines \citep[][]{jtb_trujillo01}, 
the IR triplet of Ca {\sc ii} \citep[][]{jtb_manso-trujillo-prl,jtb_manso-trujillo-10} and H$_{\alpha}$ \citep[][]{jtb_stepan-trujillo-apjl}. Moreover, it can also be used for estimating the line-center scattering polarization amplitudes expected in strong resonance lines for which frequency correlations between the incoming and outgoing photons are indeed significant (e.g., the Ly$_{\alpha}$ line investigation of Section 3.1). However, PRD effects have to be taken into account in order to model the shapes of the linear polarization profiles produced by scattering in strong resonance lines \citep[e.g.,][]{jtb_anusha}. Fortunately, the lower level of several resonance line transitions cannot carry atomic alignment because its total angular momentum is $J_l=0$ or $J_l=1/2$, so that PRD modeling efforts based on the standard two-level atom approach are of interest (e.g., the Mg {\sc ii} k line investigation considered in Section 3.2).

\section{3D modeling of scattering polarization in the quiet photosphere}

This section considers some radiative transfer investigations of scattering polarization 
in the following three-dimensional (3D) models of the quiet solar photosphere: {\bf (1)} a snapshot model taken from the hydrodynamical simulations of solar surface convection by \cite{jtb_asplund} (hereafter, the HD model) and {\bf (2)} a snapshot model taken from the magnetoconvection simulations with surface dynamo action carried out by \cite{jtb_vogler} (hereafter, the MHD model). The main aim here is to review how the modeling of observations of the intensity and linear polarization of the Sun's continuous radiation and of the Sr {\sc i} 4607 \AA\ line allows us to probe the thermal structure and the magnetization of the quiet solar photosphere. 

In both radiative transfer problems ``zero-field'' dichroism \citep[see][]{jtb_manso-trujillo-prl} can be neglected, so that the $4{\times}4$ propagation matrix of the Stokes-vector transfer equation is diagonal\footnote{For the spectral line case the propagation matrix is diagonal only if, in addition, the Zeeman splitting of $\sigma$ and $\pi$ components is negligible compared with the spectral line width. This is a good approximation for modeling the linear polarization amplitudes of the Sr {\sc i} 4607 \AA\ line observed in the quiet Sun.}. Therefore, the relevant equation for the Stokes parameter $X$ (with $X=I,Q,U$) at a given frequency $\nu$ and direction of propagation $\vec{\Omega}$ is given by

\begin{equation}
{{d}\over{d{\tau}}}{X}\,=\,{X}\,-\,S_X,  
\end{equation}
where $d{\tau}=-{\eta_I}ds$ defines the monochromatic optical distance ($\tau$) along the ray, with $s$ being the 
corresponding geometrical distance and $\eta_I$ the absorption coefficient. 
In Eq. (1) the expressions of the source-function components are

\begin{equation}
S_I=r\,S^{*}_I+(1-r)B_{\nu},
\end{equation}

\begin{equation}
S_Q=r\,S^{*}_Q,
\end{equation}

\begin{equation}
S_U=r\,S^{*}_U,
\end{equation}
where $r$ is a problem-dependent opacity ratio to be given below and  $B_{\nu}$ the Planck function. Moreover, 

\begin{eqnarray}
S^{*}_I&=\,S^0_0 + w^{(2)}_{J_uJ_l}\Big{\{}
  \frac{1}{2\sqrt{2}} (3 \mu^2-1)S^2_0 - \sqrt{3} \mu \sqrt{1-\mu^2} 
  (\cos \chi {\rm Re}[{S}^2_1] - \sin \chi {\rm Im}[{S}^2_1]) \\ \nonumber
  &\quad+ \frac{\sqrt{3}}{2} (1-\mu^2) (\cos 2\chi \, {\rm Re}[{S}^2_2]
  -\sin 2\chi \, {\rm Im}[{S}^2_2]) \Big{\}},
\end{eqnarray} 

\begin{eqnarray}
S^{*}_Q&=\, - \,w^{(2)}_{J_uJ_l}\Big{\{} \frac{3}{2\sqrt{2}}(\mu^2-1) S^2_0 -
  \sqrt{3}  \mu \sqrt{1-\mu^2} (\cos \chi
  {\rm Re}[{S}^2_1] - \sin \chi {\rm Im}[{S}^2_1]) \\ \nonumber
  &\quad- \frac{\sqrt{3}}{2} (1+\mu^2) (\cos 2\chi \, {\rm Re}[{S}^2_2]
  -\sin 2\chi \, {\rm Im}[{S}^2_2])\Big{\}}, 
\end{eqnarray}
and 

\begin{eqnarray}
S^{*}_U&=\, - \,w^{(2)}_{J_uJ_l}\sqrt{3} \Big{\{} \sqrt{1-\mu^2} ( \sin \chi
  {\rm Re}[{S}^2_1]+\cos \chi {\rm Im}[{S}^2_1]) \\ \nonumber
  &\quad+\mu (\sin 2\chi \, {\rm Re}[{S}^2_2] + \cos 2\chi \, {\rm Im}[{S}^2_2])\Big{\}},
\end{eqnarray} 
where $\theta={\rm arccos}\,({\mu})$ and $\chi$ are the inclination with respect to the solar local vertical and 
azimuth of the ray, respectively, $w^{(2)}_{J_uJ_\ell}$ is a coefficient that is equal to unity
for the problems considered in the following subsections, and the equations that govern the quantities $S^0_0$, 
$S^2_0$, ${\rm Re}[{S}^2_1]$, ${\rm Im}[{S}^2_1]$, ${\rm Re}[{S}^2_2]$ and ${\rm Im}[{S}^2_2]$ 
will be given below for each particular problem considered. We point out that in these equations the reference
direction for Stokes $Q$ is the perpendicular to the plane formed by the 
ray's propagation direction and the vertical Z-axis.

\subsection{Polarization of the Sun's continuous spectrum}

The dominant contribution to the linear polarization of the visible continuous spectrum comes from Rayleigh scattering by neutral hydrogen in its ground state (Lyman scattering) 
and Thomson scattering by free electrons \citep[][]{jtb_chandra}\footnote{The H$^{-}$ ion cannot produce polarization through Rayleigh scattering because it has a single bound state, $1s^2\,^1{\rm S}_0$. Moreover, there is no contribution from ``zero-field'' dichroism because the H$^{-}$ level cannot be polarized.}. 
The contribution of these processes to the total absorption coefficient ($\eta_I$) is quantified by ${\sigma}_c\,=\,{\sigma}_T\,N_e+\,{\sigma}_R\,n_1(H)$,
where $\sigma_T=6.653{\times}10^{-25}\,{\rm cm}^2$ is the Thomson scattering cross section, $N_e$ the electron number density, $n_1(H)$ the population of the ground level of hydrogen, and $\sigma_R$ the wavelength-dependent Rayleigh cross section. The continuum absorption coefficient is given by ${\eta_I}\,=\kappa_c\,+\,{\sigma}_c$, where $\kappa_c$ contains all the relevant non-scattering contributions (e.g., the bound-free transitions in the H$^{-}$ ion). The expressions of the $S_X$ source function components can be found easily by taking $r={{\sigma_c}\over{\kappa_c+\sigma_c}}$ and ${S^0_0}={J}^0_0$, ${S^2_0}={J}^2_0$, ${{\rm Re}[{S}^2_1]}={\rm Re}[{J}^{2}_{1}]$, ${{\rm Im}[{S}^2_1}]=-{\rm Im}[{J}{}^{2}_{1}]$, ${{\rm Re}[{S}^2_2]}={\rm Re}[{J}{}^{2}_{2}]$ and ${{\rm Im}[{S}^2_2}]=-{\rm Im}[{J}{}^{2}_{2}]$ in Eqs. (5)-(7) \citep[e.g.,][]{jtb_trujillo-shchukina-09}.

Note that the ${{J}}^K_Q$ quantities (with $K=0,2$ and $Q=0,1,2$) are the spherical components of the radiation field tensor \citep[see \S~5.11 in][]{jtb_landi-landolfi}, which quantify the symmetry properties of the radiation field at the spatial point under consideration. Thus, ${J}{}^{0}_{0}$ is the familiar mean intensity, ${J}^2_0{\approx}\oint \frac{{\rm d} \vec{\Omega}}{4\pi}\frac{1}{2\sqrt{2}} (3\mu^2-1){{I_{\nu, \vec{\Omega}}}}\,$ quantifies its anisotropy, while the real and imaginary parts of ${{J}}^2_Q$ (with $Q=1,2$) (i.e., ${\rm Re}[{J}{}^{2}_{Q}]$ and ${\rm Im}[{J}{}^{2}_{Q}]$, respectively) measure the breaking of the axial symmetry. Obviously, the real and imaginary parts of ${J}{}^{2}_{1}$ and ${J}{}^{2}_{2}$ are zero in a plane-parallel or spherically symmetric model atmosphere, but they can have significant positive and negative values in a 3D stellar atmospheric model (see figure 3 of Trujillo Bueno \& Shchukina 2009). Therefore, Eqs. (6) and (7) tell us that $U=0$ in 1D models, and that the key observational signatures of the symmetry breaking effects in a 3D model (caused by its horizontal atmospheric inhomogeneities) are non-zero $U$ signals at any on-disk position and non-zero Stokes $Q$ and $U$ signals for the LOS 
with $\mu=1$, which corresponds to forward-scattering geometry. 

\begin{figure}[!t]
\begin{center}
\includegraphics[width=15.0 cm]{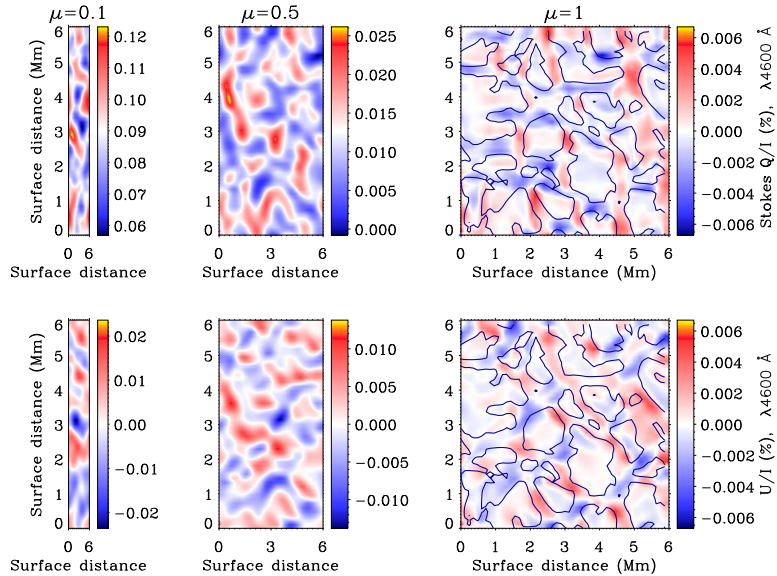}
\end{center}
\caption{The continuum scattering polarization signals at 4600 \AA\ calculated for three line of sights in Asplund et al.'s (2000) hydrodynamical model of the solar photosphere (i.e., the HD model), 
accounting for the diffraction limit effect of a 1-m telescope. 
Although the polarization amplitudes are very low at 4600 \AA, they are one order of magnitude larger at 3000 \AA. The solid line contours in the right panels delineate the upflowing granular regions. 
From \cite{jtb_trujillo-shchukina-09}.}
\label{trujillo_fig:continuum}
\end{figure}

Figure 5 shows the $Q/I$ and $U/I$ images that we would see at 4600 \AA\ if we could observe with high polarimetric precision the solar continuum polarization at very high spatial and temporal resolution. Obviously, without spatial and/or temporal resolution $U/I{\approx}0$ and the only observable quantity would be $Q/I$, whose wavelength variation at a solar disk position close to the limb has been determined semi-empirically by \cite{jtb_stenflo05}. Note that the spatially-averaged $Q/I$ of the solar continuum radiation at each $\mu$ position 
is largely determined by the effective polarizability, $r={{\sigma_c}/({\kappa_c+\sigma_c}})$, and by the fractional radiation field's anisotropy, $J^2_0/J^0_0$, which in turn depend on the density and thermal structure of the stellar atmosphere under consideration. Therefore, the more realistic the thermal and density structure of a given solar atmospheric model is, the closer to the empirical data will be the calculated linear polarization of the solar continuum radiation. Interestingly, Fig. 12 of \cite{jtb_trujillo-shchukina-09} demonstrates that 3D radiative transfer modeling of the polarization of the Sun's continuous spectrum in the above-mentioned 3D hydrodynamical model of the solar photosphere shows a notable agreement with Stenflo's (2005) semi-empirical data, significantly better than that obtained via the use of 1D semi-empirical models.

\subsection{The polarization of the Sr {\sc i} 4607 \AA\ line and the Hanle effect of a microturbulent field in a 3D hydrodynamical model of the quiet photosphere}

In this and in the following subsection we focus on the diagnostic problem of the magnetism of the quiet solar photosphere \citep[e.g., the review by][]{jtb_jos-review}. To this end, we need to measure and interpret polarization in spectral lines. It is important to note that determining the mean magnetization of the quiet Sun requires finding how much flux resides at small scales and that, to this end, it is crucial to measure and interpret the linear polarization produced by atomic level polarization and its modification by the Hanle effect.

If at the spatial grid point under consideration 
the magnetic field of strength $B$ is assumed to have an isotropic distribution of orientations within a volume ${\cal L}^3$, with ${\cal L}$ the mean-free-path of the line photons, the equations that govern the $S^K_Q$ quantities of Eqs. (5)-(7) are \citep[][]{jtb_trujillo-manso-99}:

\begin{equation}
S^0_0\,=\,(1-\epsilon){{\bar{J}}}_0^0\,+\,{\epsilon}\,B_{\nu},
\end{equation}
and

\begin{eqnarray}
  \left( \begin{array}{l} 
      {S^2_0} \\\\
      {{\rm Re}[{S}^2_1}] \\\\
      {{\rm Im}[{S}^2_1}] \\\\
      {{\rm Re}[{S}^2_2}] \\\\ 
      {{\rm Im}[{S}^2_2}]
    \end{array} \right) = w^{(2)}_{J_uJ_l} \,
  {{(1-\epsilon)} \over { 1+{\delta}^{(2)}(1-\epsilon) }}  \, 
  \left( \begin{array}{r}
      {{\bar J}}^2_0 \\\\
      {\rm Re}[{{\bar J}}{}^{2}_{1}] \\\\
      -{\rm Im}[{{\bar J}}{}^{2}_{1}] \\\\
      {\rm Re}[{{\bar J}}{}^{2}_{2}] \\\\
      -{\rm Im}[{{\bar J}}{}^{2}_{2}]
    \end{array} \right)\,{\cal {H}}^{(2)}\,,
\end{eqnarray}
where $\epsilon$ can be interpreted as the probability that an atomic excitation event is due to inelastic collisions with electrons, $\delta^{(2)}$ is the upper-level depolarizing rate (in units of the Einstein $A_{ul}$ coefficient) due to elastic collisions with neutral hydrogen atoms, and ${\cal {H}}^{(2)}$ is the Hanle depolarization factor whose value is ${\cal {H}}^{(2)}=1$ for $B=0$ G and ${\cal {H}}^{(2)}=1/5$ for $B>B_{\rm satur}$ (where $B_{\rm satur}$ is the Hanle saturation field, which is at least 200 G for the Sr {\sc i} 4607 \AA\ line). It is important to note that for the spectral line cases considered here and in the following sections the radiation field tensors are given by angular and frequency-weighted averages of the Stokes parameters, with the Voigt profile $\phi_{\nu}$ as the frequency-dependent weighting function. For this reason, in Eqs. (8) and (9) they are instead denoted with the symbol $\bar{J}^K_Q$. It is also worthwhile pointing out that for the spectral line case ${S_Q^K}\,=\,{\frac{2h{\nu}^3}{c^2}}{\frac{2{J}_l+1}{\sqrt{2{J}_u+1}}}{\rho}_Q^K$, with ${\rho}^K_Q$ being the multipolar components of the upper-level density matrix normalized to the overall population of the transition's lower level. Therefore, Eq. (9) describes 
the transfer to the atomic system of the symmetry properties of the radiation field, and note that it is the largest for line transitions with $w^{(2)}_{J_uJ_l}=1$ and when in the absence of magnetic and collisional depolarization. 
For the Sr~{\sc i} 4607 \AA\ line $J_l=0$, $J_u=1$ and $w^{(2)}_{10}=1$. Moreover, in Eqs. (2)-(4) we have now 
$r=\kappa_l\phi_{\nu}/(\kappa_l\phi_{\nu}+\kappa_c)$, with $\kappa_l$ the line-integrated opacity.

\begin{figure}[!t]
\begin{center}
\includegraphics[width=15.0 cm]{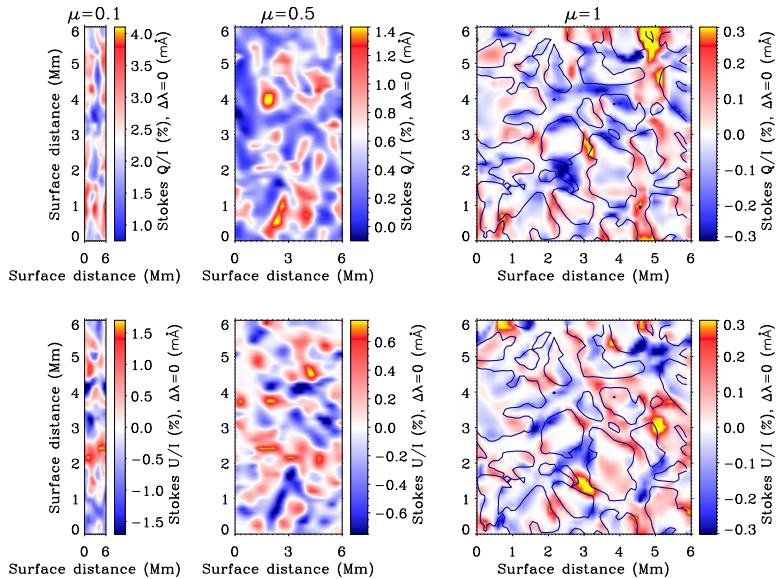}
\end{center}
\caption{The fractional linear polarization amplitudes of the Sr {\sc i} 4607 \AA\ line
calculated in the HD model of the solar photosphere, neglecting any interaction with the continuum polarization because at ${\lambda}4607$ it is almost two orders of magnitude smaller, as shown in Fig. 5. From Trujillo Bueno \& Shchukina (2007).}
\label{trujillo_fig:continuum}
\end{figure}

Figure 6 shows the center-to-limb variation of the $Q/I$ and $U/I$ line-center signals of the Sr {\sc i} 4607 \AA\ line, 
which we have obtained by solving the scattering line polarization problem in the  
above-mentioned HD model, without including any magnetic field, but accounting for the diffraction limit effect of a 1m telescope. Note that the calculated linear polarization amplitudes have sizable values and fluctuations, even at the very center of the solar disk, where we observe the forward scattering case (e.g., the standard deviations of the $Q/I$ and $U/I$ fluctuations are approximately $0.3\%$ at $\mu=0.5$). \cite{jtb_trujillo-shchukina-07} pointed out that the predicted small-scale patterns in $Q/I$ and $U/I$ are of great diagnostic value, because they are sensitive to the thermal, dynamic and magnetic structure of the quiet solar atmosphere. Therefore, it would be important to try to observe them with an imaging polarimeter attached to a suitable telescope, such as that of the SUNRISE balloon-borne telescope.

For the moment, the scattering polarization observations of the Sr {\sc i} 4607 \AA\ line that have been published lack spatial resolution (e.g., see the data points of Fig. 7). Interestingly, when the calculated $I$, $Q$ and $U$ profiles at each $\mu$ solar-disk position are spatially averaged over scales larger than those of the solar granulation pattern we obtain $U/I{\approx}0$ and the $Q/I$ vs. $\mu$ line-center amplitudes given by the green dotted line of Fig. 7. The very significant discrepancy between the observed and the calculated scattering polarization signals led \cite{jtb_trujillo-nature04} to the following conclusion: {\em the internetwork regions of the quiet 
solar photosphere are permeated by a spatially unresolved tangled magnetic field, whose 
mean field strength at heights $h{\approx}300$ km above the visible solar surface is} ${\langle B \rangle}{\approx}130$ G {\em when no distinction is made between granular and intergranular regions}\footnote{Note that 
${\langle B \rangle}{\approx}130$ G is an order of magnitude larger than the old Hanle-effect estimations by Stenflo (1982) and Faurobert-Scholl et al. (1995). For a detailed review see \cite{jtb_trujillo-spw4}.}. 
This first conclusion was reached by modeling the observed scattering line polarization through the self-consistent solution of Eqs. (8), (9) and (1) in the above-mentioned hydrodynamical model. Note that such a 3D photospheric model is unmagnetized and that, therefore, our ${\langle B \rangle}{\approx}130$ G conclusion is based on the following approximations for the description of the quiet Sun magnetic field that produces Hanle depolarization: (1) that the magnetic field is tangled at subresolution scales, with an isotropic distribution of directions and (2) that the shape of the probability distribution function, describing the fraction of quiet Sun occupied by magnetic fields of strength $B$, is exponential\footnote{It is of interest to mention that in Cattaneo's (1999) numerical experiments of surface dynamos \cite{jtb_thelen-cattaneo} noted how cleanly the PDF track an exponential behavior once the boundary effects are removed.} (i.e., ${\rm PDF}(B)={1\over{{\langle B \rangle}}}{{\rm e}^{-B/{\langle B \rangle}}}$).

\subsection{The polarization of the Sr {\sc i} 4607 \AA\ line and the Hanle effect in a 3D
magneto-convection model of the quiet photosphere with surface dynamo action}

If at the spatial grid point under consideration the magnetic field has a given strength $B$, 
inclination $\theta_B$ and azimuth $\chi_B$ (e.g., as is the case with the magnetic field of magnetoconvection simulations) the equations that govern the $S^K_Q$ quantities of Eq. (5)-(7) are \citep[][]{jtb_manso-trujillo-spw2}:

\begin{equation}
S^0_0\,=\,(1-\epsilon){\bar{J}}_0^0\,+\,{\epsilon}\,B_{\nu},
\end{equation}

\begin{eqnarray}
  \left( \begin{array}{l} 
      {S^2_0} \\\\ 
      {{\rm Re}[{S}^2_1}] \\\\
      {{\rm Im}[{S}^2_1}] \\\\ 
      {{\rm Re}[{S}^2_2}] \\\\
      {{\rm Im}[{S}^2_2}]
    \end{array} \right) = 
    w^{(2)}_{J_uJ_l} \, {{(1-\epsilon)} \over { 1+{\delta}^{(2)}(1-\epsilon) }} 
  \left( \begin{array}{r}
      {\bar J}^2_0 \\\\
      {\rm Re}[{J}{}^{2}_{1}] \\\\
      -{\rm Im}[{J}{}^{2}_{1}] \\\\
      {\rm Re}[{J}{}^{2}_{2}] \\\\
      -{\rm Im}[{J}{}^{2}_{2}]
    \end{array} \right)\,
    \end{eqnarray}
    \vspace{-0.16in}
    \begin{eqnarray}
    \hspace{1.0in} 
    -{\Gamma_u}\, {{(1-\epsilon)} \over { 1+{\delta}^{(2)}(1-\epsilon) }}\,
    \left( \begin{array}{cccccc}
        0 & M_{12} & M_{13} & 0 & 0 \\\\ 
        M_{21} & 0 & M_{23} & M_{24} & M_{25} \\\\
        M_{31} & M_{32} & 0 & M_{34} & M_{35} \\\\
        0 & M_{42} & M_{43} & 0 & M_{45} \\\\
        0 & M_{52} & M_{53} & M_{54} & 0 
      \end{array} \right) 
    \left( \begin{array}{l}
        {S^2_0} \\\\
        {{\rm Re}[{S}^2_1}] \\\\
        {{\rm Im}[{S}^2_1}] \\\\
        {{\rm Re}[{S}^2_2}] \\\\
        {{\rm Im}[{S}^2_2}]
     \end{array} \right)\,, \nonumber
\end{eqnarray}
where ${\Gamma_u}=8.79\times10^6g_{u}B/A_{ul}$ 
(with $g_{u}$ as the upper-level Land\'e factor,  
$B$ as the magnetic strength in gauss and $A_{ul}$ as the Einstein coefficient for spontaneous emission 
in ${\rm s}^{-1}$). 
The $M_{ij}$-coefficients of the magnetic kernel ${\bf M}$ 
depend on the inclination ($\theta_B$) of the local magnetic field vector 
with respect to the vertical Z-axis and on its azimuth ($\chi_B$). 
These equations for the ${S_Q^K}$ multipolar components  
have a clear physical meaning. In particular, note that the magnetic operator
${\bf M}$ couples locally the $K=2$ components among them (the Hanle effect).

\begin{figure}[!t]
\begin{center}
\includegraphics[width=9.0 cm]{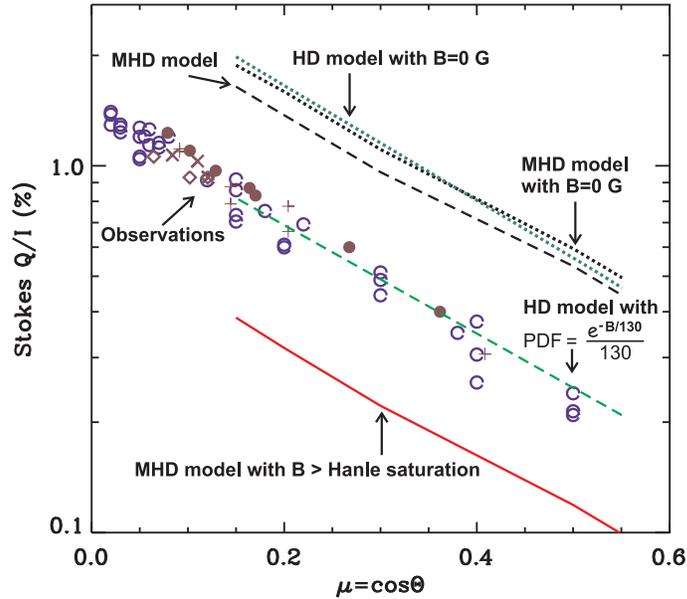}
\end{center}
\caption{Center-to-limb variation of the (spatially-averaged)  
$Q/I$ scattering amplitudes of the photospheric line of Sr {\sc i} at 4607 \AA. 
The data correspond to various observations taken by several authors during a minimum and a maximum of the solar activity cycle. The two green lines show scattering polarization amplitudes calculated 
in the 3D hydrodynamical model of \cite{jtb_asplund}, without including any magnetic field (green dotted line) and including the Hanle depolarization of a microturbulent field with an exponential PDF characterized by a mean field strength $\langle B \rangle {=} 130$\,G (green dashed line). The two black lines show scattering polarization amplitudes calculated in the MHD model of \cite{jtb_vogler}, neglecting its  magnetic field (black dotted line) and taking into account the Hanle depolarization of its magnetic field (black dashed line). We point out that the observations can be approximately fitted by multiplying each grid-point magnetic strength by a scaling factor, such that at heights $h{\approx}300$ km the resulting $\langle B \rangle$ is approximately 130 G. The red solid line shows the calculated scattering polarization amplitudes when imposing $B>B_{\rm satur}$ at each grid point in the MHD model. For more information see \cite{jtb_shchukina-trujillo}.}
\label{trujillo_fig:sun}
\end{figure}

The black dotted line of Fig. 7 shows the spatially averaged scattering polarization amplitudes of the Sr {\sc i} 4607 \AA\ line calculated by \cite{jtb_shchukina-trujillo} in the MHD model of \cite{jtb_vogler}, but after imposing $B=0$ G at each spatial grid point. Interestingly, the resulting scattering polarization amplitudes are similar (but not identical) to those computed in the HD model (see the green dotted line). The differences between the two curves are due to the fact that the thermal and density structure of the two snapshot models are not fully identical. The black dashed line shows the $Q/I$ amplitudes that \cite{jtb_shchukina-trujillo} obtain when taking into account the Hanle depolarization produced by the actual magnetic field of the MHD surface dynamo model, whose magnetic Reynolds number is $R_m{\approx}2600$. The fact that the resulting Hanle depolarization is too small to explain the observations is not surprising because the mean field strength at a height of about 300 km in the MHD snapshot model 
is only\footnote{Note also that in the idealized local dynamo experiments of \cite{jtb_cattaneo} the average unsigned field at the surface is between 10 and 20 G, if an equipartition strength of 400 G is assumed.} 
$\langle B \rangle\,{\approx}15$ G  (i.e., about an order of magnitude smaller than the $\langle B \rangle\,{\approx}130$ G value inferred by \cite{jtb_trujillo-nature04}). In order to investigate whether we can fit the observations of Fig. 7 with a magnetic field topology identical to that of the MHD surface dynamo snapshot we have increased only the magnetic strength of each spatial grid point by multiplying it by a scaling factor. Remarkably, the mean field strength that we have to reach this way to fit the observations of Fig. 7 is about 130 G \citep[see][]{jtb_shchukina-trujillo}. Note that this conclusion has now been reached without assuming that the shape of the PDF is exponential and without using the microturbulent field approximation. Clearly, both approximations were suitable to find the mean field strength of the ``hidden" field \citep[][]{jtb_trujillo-nature04}. In particular, it is interesting to note that the fact that the red line of Fig. 7 practically coincides with 1/5 of the zero-field linear polarization amplitudes indicates that the microturbulent field approximation is indeed a suitable one.

Finally, it may be of interest to mention the second conclusion of \cite{jtb_trujillo-nature04}, which resulted from their joint analysis of the Hanle effect in the Sr {\sc i} 4607 \AA\ line and in C$_2$ lines of the Swan system \citep[see also][]{jtb_trujillo-spw4}: {\em the downward-moving intergranular lane plasma is pervaded by relatively strong tangled fields at subresolution scales, with} $\langle B \rangle\,{>}B_{\rm satur}$ (where $B_{\rm satur}\,{\approx}\,200$ G is a {\em lower limit} to the Hanle saturation field of the Sr {\sc i} 4607 \AA\ line). This conclusion, which implies that most of the flux and magnetic energy reside on still unresolved scales, is also supported by the recent investigation of  \cite{jtb_shchukina-trujillo}.

\section{Scattering polarization and the Hanle effect in transition region lines}

\begin{figure}[!t]
\begin{center}
  \includegraphics[width=8.4cm]{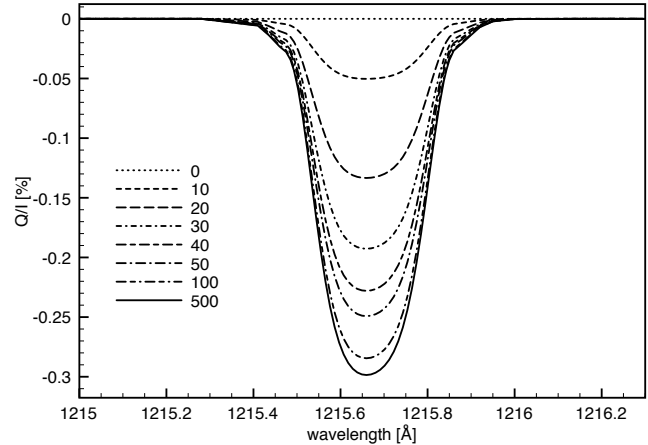}
\end{center}
\caption{Theoretical estimate of the fractional linear polarization signals of the Ly$_{\alpha}$ line
produced by forward scattering processes in the solar transition region, 
taking into account the Hanle effect caused by the
presence of a horizontal magnetic field with the indicated strength (in G). 
The reference direction for positive Stokes $Q$ is the parallel to the magnetic field vector.
Note how in forward scattering geometry (i.e., for a LOS with $\mu=1$) 
the fractional linear polarization amplitude of Ly$_{\alpha}$ {\em increases} 
as the magnetic strength of the inclined field is increased. The semi-empirical model used for the radiative transfer calculations is FAL-C \citep[see][]{jtb_falc}. From \cite{jtb_trujillo-lyman-alpha}.
}
\label{trujillo_fig:lyman-slpha}
\end{figure}

A recent review on spectropolarimetric investigations of the magnetization of the quiet Sun chromosphere can be found in \cite{jtb_trujillo-sacpeak}. In the remaining part of this section I focus instead on advancing some results of two recent theoretical investigations about the linear polarization signals expected for the hydrogen Ly$_{\alpha}$ line at 1215 \AA\ and the Mg {\sc ii} k-line at 2795 \AA, produced by scattering processes in the solar transition region plasma itself.

\subsection{Ly$_{\alpha}$ in the solar transition region} 

The intensity profile of the hydrogen Ly$_{\alpha}$ line has been
measured on the solar disk by several instruments on board rockets or
space-based telescopes (e.g., by the SUMER spectrometer on SOHO). These
measurements show that Ly$_{\alpha}$ is always in emission at all on-disk positions and times, and that the
emission originates in the upper chromosphere and transition region. The center-to-limb
variation of the observed Ly$_{\alpha}$ intensity profiles in the quiet Sun
is very small or negligible \citep[e.g.,][]{jtb_curdt}. This implies
that the {\em outgoing} radiation shows no significant variation with
the heliocentric angle $\theta$. However, this does not mean that
the illumination of the transition region atoms is isotropic -- which
would imply that scattering polarization in Ly$_{\alpha}$ is zero. The {\em incoming}
radiation (i.e., the one which illuminates the transition region atoms from ``above") 
calculated by J.~\v{S}t\v{e}p\'an (IAC) and I 
in semi-empirical models of the solar atmosphere
shows significant limb brightening, so that atoms in the transition
region are indeed irradiated by an anisotropic radiation field. The
ensuing radiation pumping induces population imbalances among the
sublevels of the upper level $2p^2{\rm P}_{3/2}$ of the Ly$_{\alpha}$ line, which
is the only hydrogen level that makes a significant contribution to its 
scattering polarization. The atomic polarization of the $2p^2{\rm P}_{3/2}$ level 
produces linear polarization in Ly$_{\alpha}$, which is
modified by the presence of a magnetic field through the Hanle effect. The $Q/I$
amplitudes we have calculated in various models of the solar atmosphere
vary between 0.5\% and  0.05\%, approximately, depending on the strength of the magnetic
field and the LOS. For example, Fig. 8 shows the calculated 
linear polarization signals caused by the Hanle effect in the forward-scattering geometry of a disk-center observation. Note that in order to be able to detect the presence of a horizontal magnetic field of 30 G in the bulk of the solar transition region the polarimetric sensitivity of the measurement should be at least 0.1\%. Interestingly, this Ly$_{\alpha}$ investigation \citep[see also][]{jtb_trujillo-cosmic} has motivated a proposal to NASA (led by Japan, USA and Spain) for the development of a Chromospheric Lyman-Alpha Spectro-Polarimeter (CLASP) for a sounding rocket \citep[see][]{jtb_CLASP}.

\subsection{Mg {\sc ii} k in the solar transition region} 

The CRD theory on which the radiative transfer modeling of the previous sections is based treats the scattering line polarization phenomenon as the temporal succession of 1st-order absorption and re-emission processes, interpreted as statistically independent events. As mentioned in the introduction, this theory  
can also be used for estimating the expected scattering polarization amplitudes in strong resonance lines for which frequency correlations between the incoming and outgoing photons are indeed significant (e.g., see Fig. 8 for the case of Ly$_{\alpha}$ in forward scattering geometry). However, it is important to emphasize that PRD effects have to be taken into account in order to be able to provide reliable predictions on the shapes of the linear polarization profiles produced by scattering in strong resonance lines like Ly$_{\alpha}$ and Mg {\sc ii} k. Recently, \cite{jtb_sampoorna-trujillo-landi-10} have studied in great detail the sensitivity of PRD scattering polarization profiles to various atmospheric parameters. To this end, the authors applied the very efficient radiative transfer method described in \cite{jtb_sampoorna-trujillo-10}. An example of the type of investigations that can be carried out with a PRD radiative transfer code based on that numerical method is considered in Fig. 9, which shows the complicated shape of the $Q/I$ profile of the Mg {\sc ii} k-line caused by scattering processes with PRD effects in a semi-empirical model of the solar chromosphere and transition region.

\begin{figure}[!t]
\begin{center}
\includegraphics[width=13.4cm]{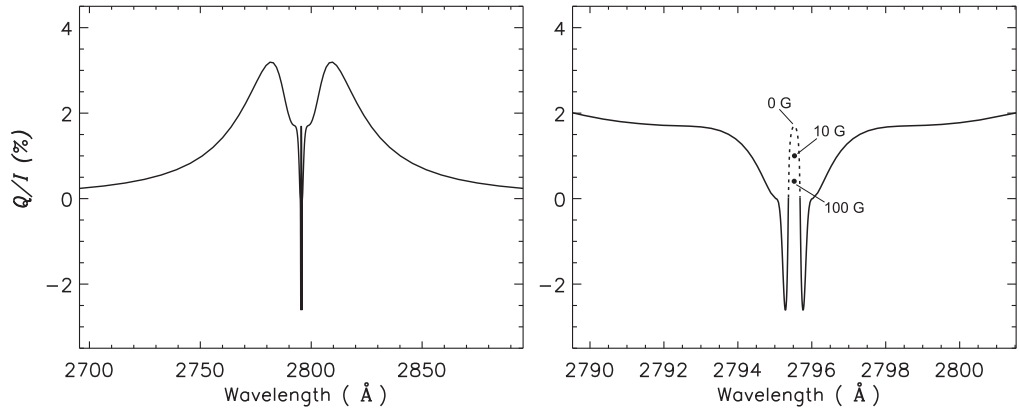}
\end{center}
\caption{The close to the limb ($\mu=0.1$) 
$Q/I$ profile of the Mg {\sc ii} k-line produced by line scattering processes in the FAL-C semi-empirical model of the solar atmosphere, taking into account PRD effects as explained in \cite{jtb_sampoorna-trujillo-10} and without accounting for the continuum polarization contribution. The right panel shows better the line-core spectral region, indicating the line-center Hanle depolarization produced by a random-azimuth horizontal magnetic field with the indicated strength. The dotted-line shows the only part of the PRD $Q/I$ profile that can be estimated via the CRD theory. The reference direction for Stokes $Q$ is as in Fig. 2.
This figure stems from a collaboration between M. Sampoorna (IIA), J.~\v{S}t\v{e}p\'an (IAC) and J. Trujillo Bueno (IAC).}
\label{trujillo_fig:lyman-slpha}
\end{figure}

Off-limb observations of spicules in Ly$_{\alpha}$ and Mg {\sc ii} k at various
heights above the visible solar limb are expected to show higher
polarization amplitudes. 
Both types of observations (on-disk and off-limb) would provide precious information on the physical
conditions of the solar transition region from the chromosphere to the
$10^6$ K solar coronal plasma, even with a spatial resolution of only 1 or 2 arcseconds.

\section{Concluding comment}

Radiative transfer modeling of solar light polarization is the essential link between theory and observations.
In general, realistic modeling requires solving a very complicated and still unsolved problem: non-LTE, multilevel, PRD radiative transfer, taking into account the joint action of the Hanle and Zeeman effects in 3D atmospheric models of the extended solar atmosphere. Fortunately, the Sun is subtle but not malicious: it allows us to learn how it works in a stepwise manner, while continuously motivating us to open up new research windows in astrophysics through the development of increasingly novel theories, instruments, modeling and data analysis techniques. 

\acknowledgments
{The results discussed here owe much to collaborations with N. Shchukina, J.~\v{S}t\v{e}p\'an, M. Sampoorna, R. Ramelli, R. Manso Sainz, M. Bianda and A. Asensio Ramos, and I thank them for fruitful discussions. I am also grateful to L. Belluzzi for carefully reviewing the paper. Finantial support by the Spanish Ministry of Science through projects AYA2010-18029 (Solar Magnetism and Astrophysical Spectropolarimetry) and CONSOLIDER INGENIO CSD2009-00038 (Molecular Astrophysics: The Herschel and Alma Era) is gratefully acknowledged.}

\end{document}